\documentclass{article}

\usepackage{ucs}
\usepackage[utf8x]{inputenc}
\usepackage[LGR,T2A,T1]{autofe}

\usepackage{graphicx}
\usepackage{amsmath}
\usepackage{url}

\usepackage{csquotes}

\usepackage{stringenc}
\usepackage[filenameencoding=utf8x]{grffile}

\usepackage{listings}
\begin{document}

\title{Transformation of Python Applications into Function-as-a-Service Deployments}
\author{Josef Spillner\\
Zurich University of Applied Sciences, School of Engineering\\
Service Prototyping Lab (blog.zhaw.ch/icclab/)\\
8401 Winterthur, Switzerland\\
josef.spillner@zhaw.ch}

\maketitle

\begin{abstract}
New cloud programming and deployment models pose challenges to software application engineers who are
looking, often in vain, for tools to automate any necessary code adaptation and transformation.
Function-as-a-Service interfaces are particular non-trivial targets when considering that most
cloud applications are implemented in non-functional languages. Among the most widely used
of these languages is Python. This starting position calls for an automated approach to transform monolithic Python code
into modular FaaS units by partially automated decomposition.
Hence, this paper introduces and evaluates Lambada, a Python module to dynamically decompose, convert and deploy
unmodified Python code into AWS Lambda functions. Beyond the tooling in the form of a measured open source
prototype implementation, the paper contributes a description of the algorithms and code rewriting rules
as blueprints for transformations of other scripting languages.
\end{abstract}

% =================================================================================================

\section{Introduction}

Software application engineers are faced with an ever-growing choice of frameworks and programming
interfaces. The issue is among the most daunting ones in contemporary service-oriented computing, namely in cloud platform services and
web frameworks \cite{webcloudtech}. The undesirable effects are \textit{continuous breakage}
and \textit{growing obsolescence}.
Instead of continuously refactoring and rewriting the application without noticeable
benefit to the end user, it is desirable to make use of automated transformations to new
target platforms \cite{autocodetransformation,syntacticrestructuring}. Apart from model-driven architectures with
targeted code generation, such solutions are also needed
for existing code where any adaptation needs to be performed directly on the code level.
This requirement applies in particular to most cloud applications which are often developed in ad-hoc ways without formal engineering processes
due to quickly evolving platform and infrastructure programming interfaces \cite{cloudsoftwareengineering}.

Application software engineering is in general a wide field divided into multiple popular design methodologies,
programming models and languages for the generation and development of code, as well as heterogeneous target platforms,
interaction patterns and runtime stacks. Many applications consist not only of code and data, but also of essential
metadata which describes how the application integrates with its environment. In the cloud applications space,
the velocity of change in the metadata space is very high. A few years back, it may have been sufficient
to describe an application in terms of a monolithic virtual machine configuration. In many recent settings, both technical
and non-technical characteristics of microservices need to be properly described, and the code and data needs
to be deployed accordingly \cite{dockerintegration}. The recent introduction of Function-as-a-Service environments
expands this path and thus presents a qualification barrier for most application engineers.

Indispensably, novel tools are needed to assist the engineers with guidance and automation to achieve live cloud service
access to their applications.
The focus of the paper is thus to explore the automated configuration, code transformation and deployment challenge specifically
for applications implemented in Python and executed as a set of hosted functions. The choice of Python for the exploration is
reasonable due to the language characteristics and its
widespread use especially for cloud application development. The work shares similarities with automated
transformation approaches for Java \cite{podilizer} but adds the perspective of dynamic code analysis and further
unique characteristics due to the transparent bytecode compilation of Python code.

To analyse the issues, the next section deduces an initial research question and justifies incremental refinements to it.
The resulting research question is then answered by first presenting the design of a transformation tool called Lambada. The design is
followed by a description of the transformation process including applicable abstract syntax tree transformation
rules. Subsequently, the implementation is described and measured and conclusions for future cloud
application development are drawn.

\section{Background and Research Question}

Python as programming language for applications has been studied in various directions.
The studies answer questions such as: how do Python programs use inheritance \cite{pythoninheritance};
how do Python programs evolve over time through code changes \cite{pythoncodechanges};
how to accelerate Python programs with cloud resources \cite{pythonacceleration};
how to offload Python program code from mobile applications to cloudlets \cite{pythonoffloading}.
Although some of the works relate to cloud infrastructure, they assume (and often introduce) specific programming interfaces
and hardware resources which are present only in prototypical form and not in commercial offerings today.
This limits their applicability to actual cloud target environments.
To the author's knowledge, there are no studies which show how to systematically transform vanilla Python applications
into deployment forms directly consumable by public cloud providers. Therefore, the general research question is
how to prepare the code and configuration of Python software applications to allow for seamless
deployment into the cloud, or in short: How are Python programs \textit{cloudified}?

As cloud computing is a wide field with various programming and deployment models, this research concentrates on hosted functions as the one which has
recently attracted a lot of attention due to cost benefits for function-sized microservices \cite{microservicecosts}
and the influx of discrete data to be processed from sensors, smart cities and connected devices \cite{cloudnativemodel}.
This Function-as-a-Service (FaaS) model gives application engineers a seemingly serverless interface.
The process of automated translation of code into deployable function units is consequently called FaaSification. The research
question then becomes more concrete: How are Python programs \textit{faasified}?

For the scoping of the work, the paper further restricts the initial research to a single management and runtime interface of FaaS.
In the absence of standards beyond single vendors, the interfaces correspond to those
implemented by AWS Lambda, while allowing flexible endpoint changes to not be limited to AWS as provider.
An additional requirement is the avoidance of any manual changes to the program source code as this could introduce
new bugs and issues. A third requirement is that Python's multi-paradigm nature is supported by translating
both functions and methods on objects into function units.
All three requirements contrast previous designs such as Briareus \cite{pythonacceleration} and PyWren \cite{pywren}.
This leads to the specific research question:

\begin{displayquote}
How are mixed-paradigm Python programs \textit{lambdafied} without requiring any modification?
\end{displayquote}

\begin{figure}[h]
\center
\includegraphics[width=\columnwidth]{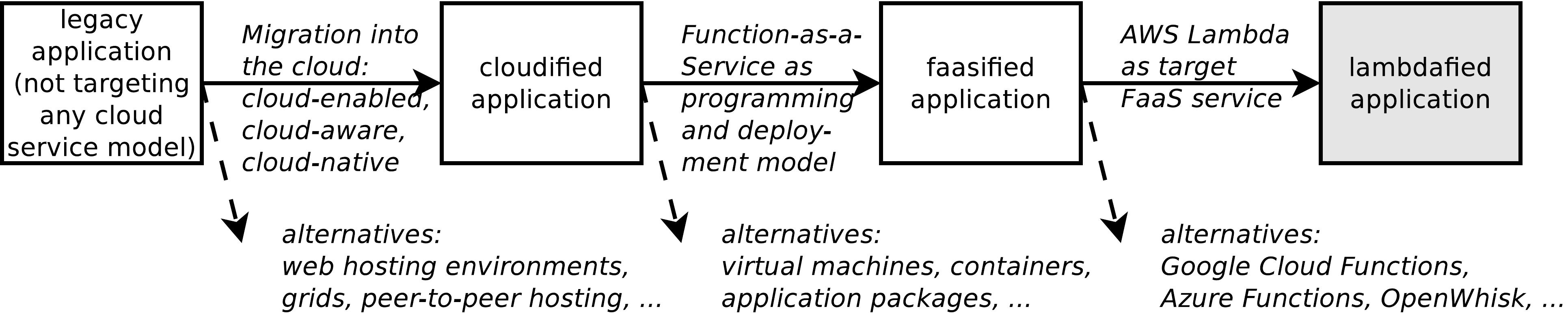}
\caption{\textit{FaaSification} and \textit{lambdafication} in context}
\label{fig:lambdafication}
\end{figure}

Fig. \ref{fig:lambdafication} puts this narrowed research scope into context. The research question is placed
on the central axis to the right side whereas possible alternative explorations are indicated as well.

\section{Tool Design}

The anticipated tool to automate the placement of Python code into Lambda is called Lambada, owing to
the ambition to turn a complex series of steps into an enjoyable activity.
Lambada is designed as a non-intrusive tool which works with existing vanilla Python code and thus targets software engineers
and developers without specialised qualifications concerning cloud computing. The principal goal is to run Python
code through Lambada in order to execute it in the scalable, resilient and invocation-centric Lambda environment.
To accomodate different use cases, the tool needs to be invocable from the command line (black box application) and from within Python
code (grey box application).
Loose coupling with the Lambda management (control plane) interface is achieved by not interfacing directly through
a corresponding library (e.g. Boto), but instead relying on the default command-line utility (AWS CLI) including
prior region and credentials configuration.

The Lambada tool dynamically imports the application code, inspects the namespace, and transforms functions and
classes into local proxies paired with generated remote functions. Referenced module imports are inspected and their
contents are transformed recursively. The subsequent processing depends on the chosen mode. In production mode, all remote functions are deployed
upon first invocation and are later used directly, suggesting a slowdown for the first invocation. The main code
block of the application is then executed with references to the remote functions. In debug mode, the generated
and rewritten functions are instead serialised into local files and no execution takes place.

The design of the tool is influenced by the expected inputs and outputs. The inputs are assumed to be
moderately complex Python applications consisting of a primary module and its dependency modules
which contain mixed amounts of procedural (\texttt{def}) and object-oriented (\texttt{class}/\texttt{def}) code.
All methods and functions have arbitrary parameter counts and types.
The outputs are functions adhering to the Lambda signature of two parameters \texttt{event} and \texttt{context}
in a function by default named \texttt{lambda\_handler}.

\begin{figure}[h]
\center
\includegraphics[width=0.75\columnwidth]{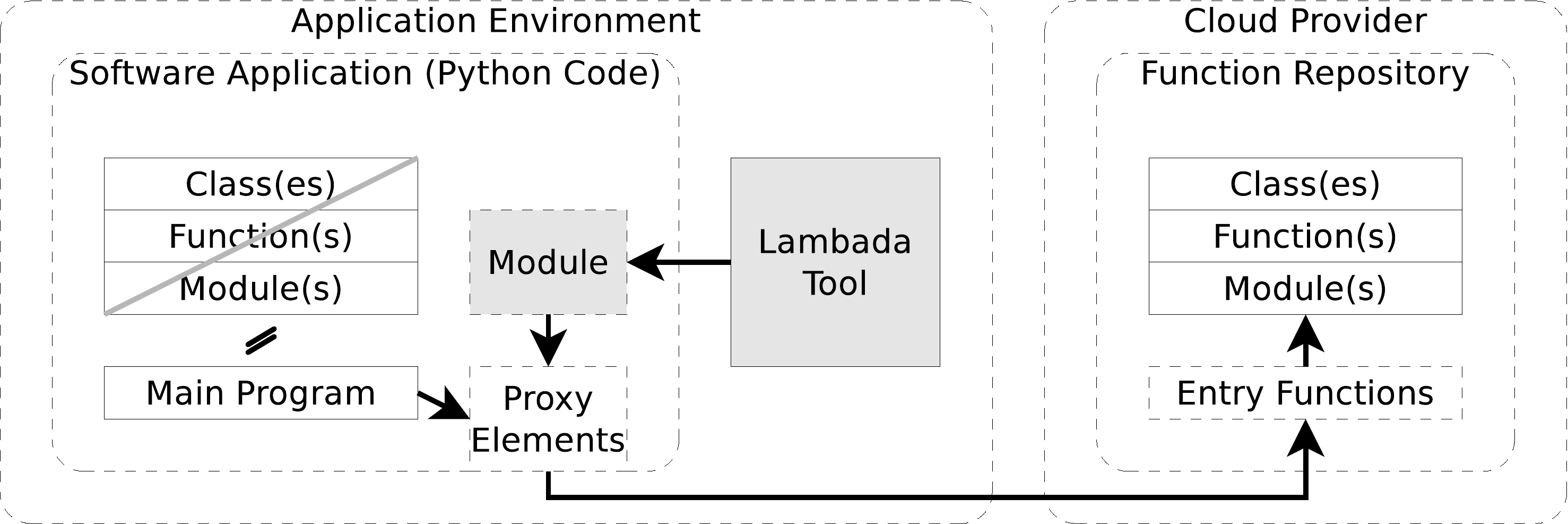}
\caption{Design of the Lambada code transformator}
\label{fig:lambadadesign}
\end{figure}

Fig. \ref{fig:lambadadesign} shows the anticipated tool design. The dashed parts are generated and/or injected
into the application at runtime. The striked parts are deleted from the interpreter memory and thus removed from the execution.

The code transformation is based on function and class definitions on the input side. Each function and method is
projected onto a single hosted function. While the code introspection happens dynamically, it is subject to
static feasibility checks. Functions and methods containing legitimate bodies are admitted to the transformation
independently from foreseeable execution issues due to restrictions in the target environment. In AWS Lambda and
many competitor services, these restriction encompass both time (e.g. 300 seconds maximum duration) and space (e.g. 1.5 GB maximum
allocation of RAM).

\begin{figure}[h]
\center
\includegraphics[width=\columnwidth]{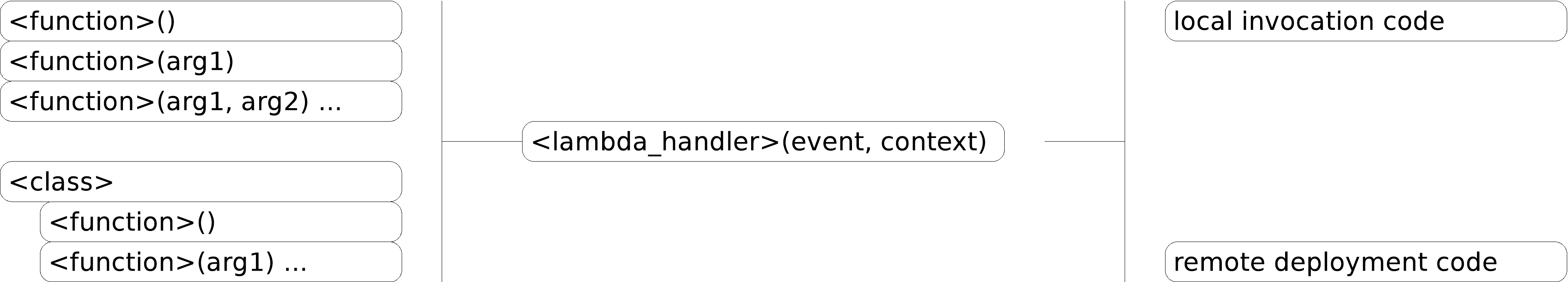}
\caption{Paths in the Lambada code transformator}
\label{fig:lambadatrafo}
\end{figure}

Lambada makes use of two output channels: Persistent local files for rewritten code, corresponding to debug mode, and ephemeral function unit files which
are partially deployed into Lambda, corresponding to production mode.
Fig. \ref{fig:lambadatrafo} shows the paths for transforming function code including the input and output elements.

\section{Transformation Process}

Code transformation may refer to syntax changes on different levels. In CPython, the default interpreter for Python,
source code is first tokenised (token/parser level), then assembled into an abstract syntax tree (tree level), and finally bytecode-compiled.
Although other interpreters exist \cite{pylibjit}, access to the syntax tree is always possible at runtime.
The code transformation in Lambada is performed with specific rules for Python modules, functions, classes and methods
on the syntactic level. It also considers standard input and output adaptations to the hosted functions environment.
There are two kinds of rules: direct manipulations of the abstract syntax tree, and code templates which
are filled, compiled and inserted at appropriate nodes of the tree.

The rules apply to five major syntax elements: modules, functions, classes with methods, globals as well as
called built-ins for standard input and output. These elements are described in the next paragraphs.

\subsection{Modules}

In Python, each script file represents a module which is loaded with the file name but omitting the \texttt{.py}
suffix. To \textit{lambdafy} an application, the set of modules belonging to this application needs to be determined.
This can be accomplished on the file level by checking which files belong to a certain application project,
or on the function level by recursive traversal of dependency functions. The second approach is more precise
as it does not attempt to transform files with code paths unreachable from a certain function, but also
requires more care to stop the traversal for modules outside of the application scope such as system modules.
Another advantage is that it allows for dynamic analysis so that modules which are imported conditionally,
or based on evaluated code, will be identified correctly.
Lambada thus follows the second approach.

\begin{figure}[h]
\center
\includegraphics[width=0.8\columnwidth]{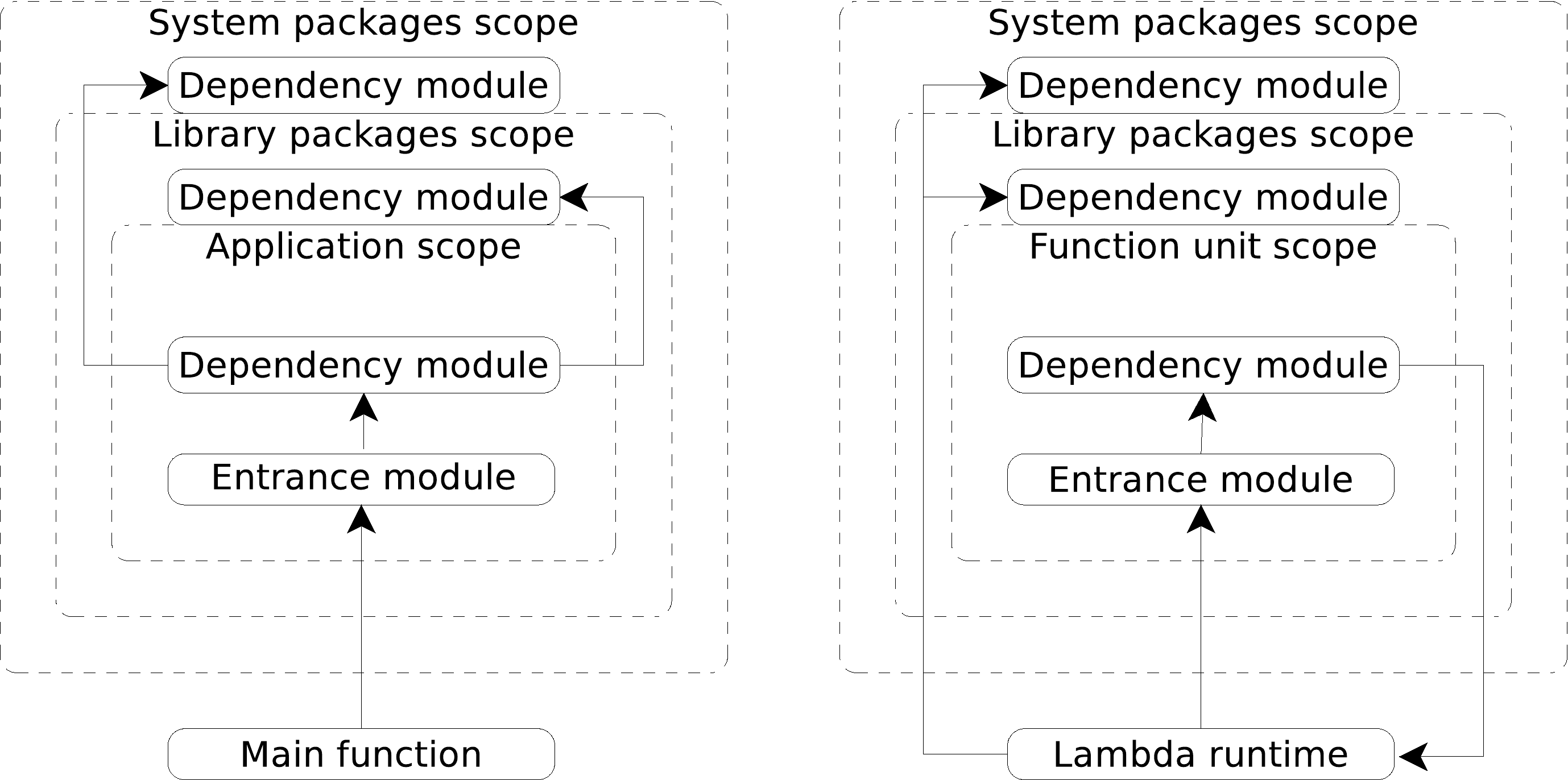}
\caption{Function dependency analysis across module boundaries}
\label{fig:modules}
\end{figure}

Fig. \ref{fig:modules} gives an example of the traversal. On the left side, an application consists of code which
references a module called entrance module. Multiple dependency modules are referenced from it, of which only
some belong into the application scope. On the right side, the transformed code is represented. As all functions
are transformed into corresponding Lambda units, their invocations across module boundaries involve the Lambda
runtime as gateway.

\subsection{Functions}

The abstract syntax tree is recursively walked. Interceptions happen at function invocations (calls)
and function definitions. A dependency map between calling and called functions is continuously updated
at each call interception. The function analysis does not encompass inline anonymous functions which
in Python are, ironically, called Lambdas.

Code templates exist for remote function equivalents, local call stubs and local replacements, as well
as for function proxies for all dependencies. They are instantiated and filled for each definition whereas
calls are merely rewritten to point to the call stubs instead. Furthermore, the remote function units
are packaged along with suitable configuration into archive files and deployed into the target Lambda
environment.

\begin{figure}[h]
\center
\includegraphics[width=0.69\columnwidth]{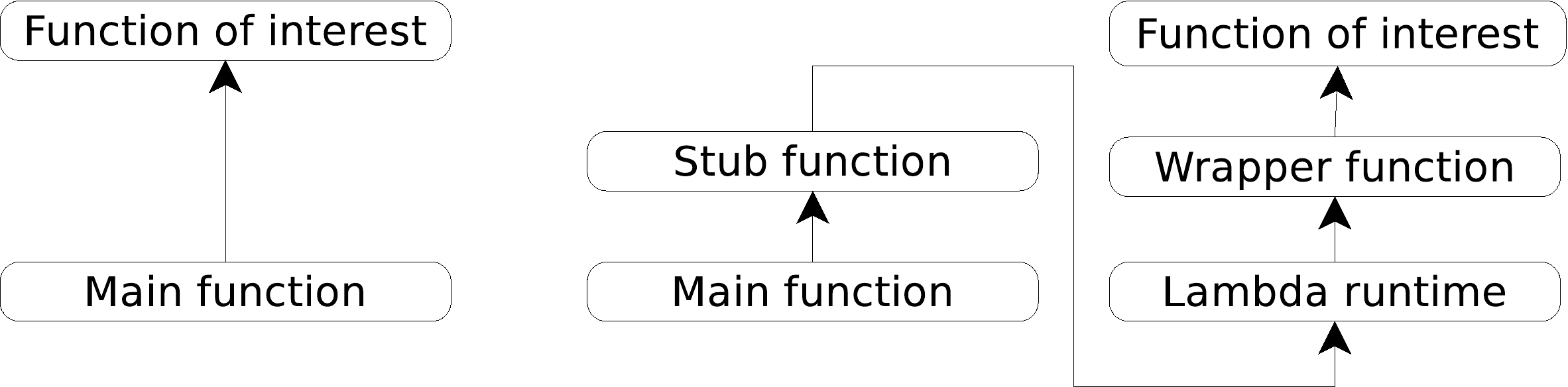}
\caption{Transformation of functions}
\label{fig:functions}
\end{figure}

Fig. \ref{fig:functions} gives an example of a function transformation. The original function which
had a single local entrypoint is replicated unmodified or with slight modifications and placed into the
target environment which leads to a second entrypoint. A local stub and a remote skeleton as parameter-compliant
wrapper function connect both environments.

\subsection{Classes}

The programming model of FaaS mandates functions. Classes therefore need to be decomposed into their methods which
are then represented as stateless functions, receiving the object state for each invocation as additional parameter.
One technique to do so is name mangling as known from linking C++ binary code \cite{mangling}.
Another technique is method call interception. It fits well into the anticipated generic programming model and is
therefore the chosen technique which will be presented.

Two generic proxy classes, one on the service side and one on the client side, perform the exchange of
function names, call arguments and associated object state. The client-side proxy is a metaclass which
is instantiated into a derived class with a hidden attribute referring to the class to be replaced. Objects
of this class then accept all attribute and method accessors as the substituted class.
On the client side, all class definitions are dynamically overwritten with this proxy class.

On the service side, all class definitions are used almost unmodified. The only necessary change is the introduction of
a separate constructor method (\texttt{\_\_remote\_\_init\_\_}) which can be intercepted similar to regular methods, and
its invocation from the original constructor (\texttt{\_\_init\_\_}).

Method calls transmit the module name, class name, method arguments and a dictionary representation of all attributes.

Due to the limitation imposed by most FaaS models to only allow for a single entry-point function per file or set of files,
the class definition needs to be duplicated as separate hosted function for each method.

\subsection{Globals}

In contrast to most compiled programming languages, the syntax of Python does not foresee a main function.
The language's equivalent of a main function in a module is the ordered set of all lines not part of a function
or class definition. By convention, a stricter equivalent is marked as a conditional branch whose condition
reads \texttt{if \_\_name\_\_ == "\_\_main\_\_"} and whose code block contains the function implementation.
To make Lambada practical, such blocks are replicated verbatim in the local rewritten file.

Furthermore, global variables may be referenced by functions. While considered bad practice, this pattern is still
common practice and hence is targeted by the transformation by placing a replicated global variable alongside each
affected deployed function. The use of such variables to synchronise state, for instance through message queues, is therefore
inhibited implicitly. Such changes in programming semantics can only be discovered through subsequent unit testing.

\subsection{Input and Output}

Lambada scans the syntax tree for any occurence of \texttt{input} or \texttt{print} (output) function calls per function.
If the presence is detected, monads are set up to carry any data input and output across invocations and across the network
to and from the caller. The print monad is a global variable which concatenates all text and finally returns it as second
return value next to the actual one to the caller.

\section{Implementation and Evaluation}

Lambada is implemented as Python 3 executable module following the proposed design.
The code size is around 300 lines for handling functions and their transformation and deployment and 100 lines for handling classes and methods,
plus 20 lines for the application wrapping the module. These numbers exclude external modules for code generation (codegen)
and any battery modules used from the standard Python distribution.
In the following, selected implementation details are presented before proceeding to the
presentation of evaluation results.

\subsection{Proxy Classes}

Listing \ref{lst:app} describes the implementation of the proxy class for transparent lambdafication of methods.
Python's \texttt{\_\_new\_\_} method is therein called twice: first, to return an object of the proxy class with
a flag indicating it, and second, an instance of the actual object wrapped into the proxy object. Subsequently,
any method or attribute access is handled by the proxy which transfers state between the local object and the
lambda-side object, keeping both in sync for any potentially stateful method invocation.

\begin{lstlisting}[caption=Proxy class implementation,label=lst:app,mathescape]
class Proxy:
	def __new__(cls, classname, proxy=True):
		if proxy:
			return lambda: Proxy(classname, False)
		else:
			return object.__new__(cls)

	def __init__(self, classname, ignoreproxy):
		self.classname = classname
		self.__remote__init__()

	def __getattr__(self, name):
		def method(*args):
			cn = self.classname
			_d = json.dumps(self.__dict__)
			_dc = json.loads(_d)
			del _dc["classname"]
			_d = json.dumps(_dc)
			_d, *args = netproxy.Netproxy(_d, self.classname, name, args)
			self.__dict__ = json.loads(_d)
			self.__dict__["classname"] = cn
			return args
		return method
\end{lstlisting}

\subsection{Invocation}

There are three different ways to invoke Lambada: as a command-line application, as an executable Python module,
and as a regular Python module.
For integration with development environments, for instance as part of continuous testing, integration and deployment chains,
Lambada runs as standalone application targeting any files specified as command-line parameters.
Only in this way Python scripts containing functions and methods can be rewritten locally.
The pre-loaded module is a main
program which is run before the application main program and hooks into all globals such as imported modules and function declarations.
Being a module itself, Lambada can also be imported
into the application to consciously and selectively transform parts of the application in case minimal code modifications are permitted.

In the first two cases,
the target module is specified, whereas in the third case, the current execution context is assumed as starting
point from which, in all three cases, imported modules are recursively transformed when demanded by function
dependencies not part of the standard library which is also expected in the target environment.
Furthermore, in all three cases, a custom endpoint can be specified in case the functions are intended to be
uploaded not to AWS Lambda but to reimplementations thereof, such as Snake Functions \cite{snafu}.

\subsection{Experimental Evaluation}

The practical use of Lambada is evaluated empirically by transforming self-contained Python functions and applications
and measuring both the transformation overhead and the resulting execution overhead.
All results have been obtained on a Notebook running Linux 4.9.0 and Python 3.5.3 on an Intel i7-5600U CPU with 4 cores à 2.60 GHz.
The network connection to AWS Lambda's US-West-1 region is located on SWITCHlan, the 100 GBit/s Swiss research and education network, with an ICMP roundtrip
time of 2 ms, jitter of 3 ms and effective transfer rates of 94 Mbps averaged during the experimental runs.
All Lambda instances have been configured as smallest with 128 MB of RAM which inherently influences the execution speed through
an opaque mapping of memory allocation to CPU allocation within the service.
The material to reproduce the experiment under similar or different conditions is available along with the Lambada implementation.

Table \ref{tab:fibperf0} describes a purely local example of a single recursive Fibonacci function \texttt{fib} implemented in Python.
The original implementation is lambdafied, rewritten to a local file copy, and locally executed with Lambda execution
syntax and semantics including JSON serialisation. The function instance is parameterised with a value $x$, leading to
a result of $y = fib(x)$ involving $2y-1$ total function calls. The overhead per function invocation contains the constant
time of $L = 67ms$ for the lambdafication process itself and is calculated as per Eq. \ref{eq:overhead}.

\begin{table}[htb]
\centering
\caption{Local Fibonacci comparison.\label{tab:fibperf0}}
\begin{tabular}{|l|r|r|r|r|} \hline
\textbf{Invocation}	& \textbf{Calls}	& \textbf{Original}	& \textbf{Lambdafied}	& \textbf{Overhead}	\\ \hline

fib(1)			& 1			& 0.0003 ms		& 0.0210 ms		& 223402.33		\\ \hline
fib(10)			& 109			& 0.0202 ms		& 1.3200 ms		& 396.03		\\ \hline
fib(20)			& 13529			& 1.57 ms		& 156.47 ms		& 98.67			\\ \hline
fib(30)			& 1664079		& 191.81 ms		& 19639.70 ms		& 101.39		\\ \hline
fib(40)			& 204668309		& 24989.58 ms		& 2382736.84 ms		& 94.34			\\ \hline
\end{tabular}
\end{table}

\begin{equation}
\label{eq:overhead}
overhead = \frac{T_{lambda} + \frac{L}{(2y-1)}}{T_{orig}} - 1.
\end{equation}

The lambdafication overhead is extremely significant for functions which are invoked once and insignificant for functions
invoked thousands of times. The lambdafied execution overhead is still highly significant independent of the number of
invocations with a slowdown factor of around 100 mostly due to JSON parsing.
It should be noted that by using the optimised Python interpreter PyPy instead of the
standard CPython implementation, the absolute execution times are reduced to a varying level of around 45-70\%
and the overhead factor drops from around 100, where it stabilises for higher values of $x$, to around 67. These limits are representing the practical upper boundary of overhead
as most functions would contain a higher share of computation.
Another interesting observation is that \texttt{fib(30)} would already exceed the free tier limit of 1 million
function invocations at AWS Lambda.
For completeness, the function source code is shown in Listing \ref{lst:fib}.

\begin{lstlisting}[caption=Fibonacci function implementation,label=lst:fib,mathescape]
def fib(x):
	if x in (1, 2):
		return 1
	return fib(x - 1) + fib(x - 2)
\end{lstlisting}

The second such function is an extended recursive Fibonacci function \texttt{fibs} which on each invocation calculates the
invocation count's number of sine values. This makes it more compute-intensive and less invocation-intensive,
avoiding high experimentation cost while now including the deployment to and runtime at AWS Lambda as well.
It is also a test case for handling global variables since the invocation count needs to be maintained persistently across
invocations.
The overhead per function invocation therefore includes two constants, the lambdafication process time (again $L = 67ms$)
and the deployment time (network-dependent, around $D = 4200ms$ on the test system), leading to a shared overhead part
of $\frac{L + D}{(2y-1)}$.
Table \ref{tab:fibperf} compares the performance values
for selected parameters averaged over 100 command-line invocations each. Due to the higher computation complexity,
the parameter $x$ is set to lower values compared to the previous run, although \texttt{fibs(1)} and \texttt{fibs(10)} are contained in both sets.
Already with \texttt{fibs(19)} the Lambda instance will time out.

\begin{table}[htb]
\centering
\caption{Local and Lambda Fibonacci comparison.\label{tab:fibperf}}
\begin{tabular}{|l|r|r|r|r|} \hline
\textbf{Invocation}	& \textbf{Calls}	& \textbf{Original}	& \textbf{Lambda}	& \textbf{Overhead}	\\ \hline

fibs(1)			& 1			& 0.04 ms		& 18.94 ms		& 107147.49		\\ \hline
fibs(10)		& 109			& 1.33 ms		& 1157.77 ms		& 898.94		\\ \hline
fibs(12)		& 287			& 6.54 ms		& 7041.89 ms		& 1078.01		\\ \hline
fibs(15)		& 1219			& 117.31 ms		& 14857.44 ms		& 125.68		\\ \hline
fibs(18)		& 5167			& 2235.03 ms		& 63893.69 ms		& 27.59			\\ \hline
\end{tabular}
\end{table}

The observations are remarkable. First, there is a slowdown factor in the order of magnitude of significantly less than 100 for
invocations with longer compute periods despite including the network deployment. This makes a number of real-world
use cases without strict latency requirements possible.
Second, the invocation time variance is much higher with Lambda. This suggests that performance-critical
code sections will benefit from typical optimisation techniques such as inner functions or local function calls
at the expense of more code duplication.

\section{Discussion}

FaaSification is an emerging technique whose application fields need to be determined. Two interesting and feasible ones
are the transformation of (admittedly simple) legacy applications and the continuous local testing of applications
under development. The distinct methods include so far both static and dynamic decomposition of modules into functions,
but not yet functions into smaller functions to avoid hitting deployment limits in the target services described in this
paper. New concepts are needed to manage the resulting tiny microservices, or nanoservices. Current management systems,
including AWS Lambda and the related AWS API Gateway, are evidently optimised to handle few higher-value functions instead of massive
amounts of smaller functions. The elimination of duplicated dependency code and the optimal selection of function instance sizes
due to their influence on execution performance will be other targets for future research.

Even with the prototypical nature of Lambada, the tool already proves useful in programming education beyond computer
science and engineering curricula. Students often struggle with formulating their algorithms. Deploying functions individually
in a manual process would introduce a chilling effect and reduce the motivation to focus on the core problems significantly.
Automated decomposition and deployment assist to maintain the motivation.

\section{Conclusion}

Automated code deployment and transparent code offloading to FaaS are interesting new
workflows in cloud application software engineering and testing scenarios. This paper has introduced
and explained Lambada, a tool to automate this workflow. The tool shifts Python functions and methods along
with module-internal code dependencies into hosted function services implementing the AWS Lambda management
interface. While the overheads are significant, they are determined mostly by constants whose effect is reduced
in more complex function with a runtime of more than a few seconds. This makes the approach feasible for
a number of development and testing tasks regularly performed by cloud application engineers.

\section*{Repeatability}

Lambada is publicly available at \url{https://gitlab.com/josefspillner/lambada}.
The repository contains the Fibonacci functions references in this paper among other examples.
To repeat the experiments, a starting point is the invocation of \texttt{./lambada --debug --local examples/fib.py}
which rewrites the script containing the function locally for inspection and informs about the transformation process.
Omitting the flag \texttt{--debug} is recommended for measurements. Omitting \texttt{--local} furthermore
performs the deployment and execution in the Lambda environment assuming the AWS CLI tools are installed, configured
and working correctly.

\section*{Acknowledgements}

This research has been supported by an AWS in Education Research Grant which helped us to run our experiments on AWS Lambda as representative public commercial FaaS.

\bibliographystyle{unsrt}
\bibliography{lambadapaper}

\end{document}